\documentclass[twocolumn]{aastex63}

\usepackage{amsmath}
\usepackage{algorithmic}

\received{2020 August 28}
\revised{2020 October 19}
\accepted{2020 November 10 at ApJS}

\turnoffedit

\begin{document}

\allowdisplaybreaks

\title{Analysis of coronal mass ejection flux rope signatures using 3DCORE and approximate Bayesian Computation}

\correspondingauthor{Andreas J. Weiss}
\email{andreas.weiss@oeaw.ac.at}

\author{Andreas J. Weiss}
\affiliation{Space Research Institute, Austrian Academy of Sciences, Schmiedlstraße 6, 8042 Graz, Austria}
\affiliation{Institute of Physics, University of Graz, Universit\"atsplatz 5, 8010 Graz, Austria}
\affiliation{Institute of Geodesy, Graz University of Technology, Steyrergasse 30, 8010 Graz, Austria}

\author{Christian M\"ostl}
\affiliation{Space Research Institute, Austrian Academy of Sciences, Schmiedlstraße 6, 8042 Graz, Austria}
\affiliation{Institute of Geodesy, Graz University of Technology, Steyrergasse 30, 8010 Graz, Austria}

\author{Tanja Amerstorfer}
\affiliation{Space Research Institute, Austrian Academy of Sciences, Schmiedlstraße 6, 8042 Graz, Austria}

\author{Rachel L. Bailey}
\affiliation{Space Research Institute, Austrian Academy of Sciences, Schmiedlstraße 6, 8042 Graz, Austria}
\affil{Zentralanstalt f\"ur Meteorologie und Geodynamik, Hohe Warte 38, 1190 Vienna, Austria}

\author{Martin A. Reiss}
\affiliation{Space Research Institute, Austrian Academy of Sciences, Schmiedlstraße 6, 8042 Graz, Austria}
\affiliation{Institute of Geodesy, Graz University of Technology, Steyrergasse 30, 8010 Graz, Austria}

\author{J\"urgen Hinterreiter}
\affiliation{Space Research Institute, Austrian Academy of Sciences, Schmiedlstraße 6, 8042 Graz, Austria}
\affil{Institute of Physics, University of Graz, Universit\"atsplatz 5, 8010 Graz, Austria}

\author{Ute A. Amerstorfer}
\affiliation{Space Research Institute, Austrian Academy of Sciences, Schmiedlstraße 6, 8042 Graz, Austria}

\author{Maike Bauer}
\affiliation{Space Research Institute, Austrian Academy of Sciences, Schmiedlstraße 6, 8042 Graz, Austria}
\affil{Institute of Physics, University of Graz, Universit\"atsplatz 5, 8010 Graz, Austria}

\begin{abstract}
We present a major update to the 3D coronal rope ejection (3DCORE) technique for modeling coronal mass ejection flux ropes in conjunction with an Approximate Bayesian Computation (ABC) algorithm that is used for fitting the model to in situ magnetic field measurements. The model assumes an empirically motivated torus-like flux rope structure that expands self-similarly within the heliosphere, is influenced by a simplified interaction with the solar wind environment, and carries along an embedded analytical magnetic field. The improved 3DCORE implementation allows us to generate extremely large ensemble simulations which we then use to find global best-fit model parameters using an ABC sequential Monte Carlo (SMC) algorithm. The usage of this algorithm, under some basic assumptions on the uncertainty of the magnetic field measurements, allows us to furthermore generate estimates on the uncertainty of model parameters using only a single in situ observation. We apply our model to synthetically generated measurements to prove the validity of our implementation for the fitting procedure. \edit1{We also present a brief analysis, within the scope of our model, of an event captured by Parker Solar Probe (PSP) shortly after its first fly-by of the Sun on 2018 November 12 at 0.25 AU. The presented toolset is also easily extendable to the analysis of events captured by multiple spacecraft and will therefore facilitate future multi-point studies.}
\end{abstract}


\section{Introduction} \label{sec:intro}

Coronal mass ejections (CMEs) are the most violent and energetic events that occur within our solar system and have a significant impact on the interplanetary magnetic field and planetary magnetospheres \citep{Schwenn_2005, Chen_2011, Webb_2012}. An enormous amount of magnetized plasma is ejected into the interplanetary medium and propagates as an extremely large, and continuously expanding, structure \citep[e.g.][]{Burlaga_1981,Farrugia_1993,Gopalswamy_2000} that can reach the outer planets of our solar system \citep[e.g.][]{Witasse_2017}. CMEs also carry along a strong internal magnetic field, believed to be in the form of a magnetic flux rope \citep[MFR, e.g.][]{Marubashi_1986, Burlaga_1988, Lepping_1990}, that can manifest itself as a magnetic cloud when measured in situ \citep{Burlaga_1981, Klein_1982, Bothmer_1988} by a spacecraft. This strong magnetic field, given a certain configuration of the CME, can also induce what is known as a geomagnetic storm  \citep[e.g.][]{Farrugia_1993, Gonzalez_1994}. Geomagnetic storms are associated with a variety of phenomena such as aurorae, geomagnetically induced currents \citep[e.g.][]{Pirjola_1983, Boteler_1998, Bolduc_2002}, and disturbances within the ionosphere \citep[e.g.][]{Proelss_1980}, which adversely affect high-frequency ground to ground radio or spacecraft communication. CMEs, alongside other solar events, can furthermore pose a significant radiation hazard for human space travel \citep[e.g.][]{Zeitlin_2013}.

As such, the study of CMEs has been of high interest to the solar physics and space weather community ever since their discovery during the Skylab era \citep{Tousey_1973} and connection with terrestrial phenomena \citep{Gosling_1991}. Nonetheless, there remain a number of unresolved issues regarding their generation, interplanetary evolution and internal structure. Some of these issues are exacerbated due to the fact that most CMEs are only observed by individual satellites and their global structure therefore remains largely hidden. Solar imagers and coronagraphs are capable of showing the formation and eruption phases of CMEs but these observations are susceptible to projection effects. They are also not fully representative of the global structure within the interplanetary medium as CMEs can undergo drastic change due to rotation or deflection \citep[e.g.][]{Vourlidas_2011, Kay_2015, Moestl_2015} or interaction with the solar wind \citep[e.g.][]{Riley_2004, Manchester_2017, Luhmann_2020}.

Further uncertainty exists on the structure of the internal magnetic field, which determines the geoeffectivity. \edit1{The magnetic field structures have been approximately described using several different flux rope models} such as those based on the analytical cylindrical Lunquist solution \edit1{\citep{Lundquist_1950, Burlaga_1988, Zhang_1988, Lepping_1990, Owens_2006A, Owens_2006B, Kay_2017}} or the analytical cylindrical Gold-Hoyle solutions \citep[e.g.][]{Gold_1960, Farrugia_1999}. The most recent studies have also introduced models for slightly altered geometries such as torii \citep[e.g.][]{Hidalgo_2012, Vandas_2017}  or elliptical cylinders \citep[e.g.][]{Hidalgo_2002, Nieves_Chinchilla_2018} in order to better accommodate for distortions further away from the idealized cylindrical flux rope picture. Other models that do not use flux ropes include those that are based on spheromaks \citep{Farrugia_1995, Vandas_1997}. \edit1{Additionally,} non-specified field structures were calculated with the Grad-Shafranov equation \citep{Hu_2002, Moestl_2009}, and they result in a field closely resembling a Gold-Hoyle flux rope \citep{Hu_2015}. In many cases multiple different models are able to reproduce the same observed measurements with similar accuracy and are therefore indistinguishable in the absence of independent auxiliary measurements.

A better understanding of the global structure of CMEs and their embedded magnetic field can be gained by observing multi-point events, i.e. CMEs that were observed by multiple spacecraft at distinct positions within the heliosphere. These events are inherently rare due to the low number of operating spacecraft and only a few dozen have been cataloged so far \citep[e.g.][]{Burlaga_1981, leitner_2007, Good_2019, Vrsnak_2019, Salman_2020}. With the advent of next-generation spacecraft dedicated to solar physics such as Parker Solar Probe (PSP) \citep{Fox_2016} or Solar Orbiter \citep{Mueller_2013}, along with other missions such as Bepi Colombo that have onboard magnetometers, longer operating spacecraft such as STEREO-A \citep{kaiser_2008} as well as possible interplanetary CubeSats, there will be up to half a dozen spacecraft within 1~AU sampling the heliosphere in the upcoming years. Combined with the rise of solar cycle 25, there will likely be a considerable number of additional multi-point CME observations available.


\edit1{
The subsequent study of such multi-point events will require global CME models that can simulate measurements at several distinct positions simultaneously. The general approach would require expensive MHD simulations of the inner heliosphere for the solar wind environment and the evolving CME. Examples of such simulation codes include Enlil \citep{Odstrcil_2004}, EUHFORIA \citep{Pomoell_2018, Verbeke_2019} or MAS \citep{Toeroek_2018}. Unfortunately, the use of these models is limited due to their computational complexity and inherent issues in the generation of the boundary conditions on and near the Sun. On the other hand, the most simple analytical flux rope models that are used for describing the magnetic field locally \citep{Lepping_1990, leitner_2007} make use of rigid geometries that are incapable of accounting for any of the interplanetary evolution for CME structure that is expected to occur. Due to these obstacles, there has been a recent focus on using semi-empirical models as an alternative \citep{Isavnin_2016, Moestl_2018, Rouillard_2020}. Semi-empirical models have the major advantage that they are computationally inexpensive and conceptually easy to implement. It is furthermore possible, with varying degrees of complexity, to include simplified interactions with the solar background wind. These models should be seen as an attempt to bridge the gap between the ``expensive'' MHD simulations and the geometrically simple but physically accurate analytical models.
}

In this paper we will showcase a significant update to the 3D coronal rope ejection (3DCORE) model \citep{Moestl_2018} which is such a semi-empirical model. 3DCORE is a forward simulation model that describes a CME as a propagating and \edit1{self-similarly} expanding torus-like structure, influenced by a simple drag model that stays attached to the Sun. The cross-section of this torus is extended to also incorporate elliptical shapes (i.e. different aspect ratios). This torus-like structure contains an embedded Gold-Hoyle-like field with a time-invariant twist number.

\edit1{
Our assumption of self-similar expansion for a torus-like structure is not able to describe deformations of the flux rope structure that can occur due to drastic longitudinal solar wind speed gradients. Nonetheless, it can be expected that the approximation is generally adequate for modeling measurements on a small scale. The model can therefore also be used to infer the distance scales at which the approximation breaks down. This can be of use for future studies as it gives an estimate for the required spatial resolution for resolving deformations.
}

We improve the \edit1{3DCORE} model implementation \edit1{so that it} allows us to generate extremely large \edit1{mega-}ensembles on the order of $10^6$ runs per second. We use this newly gained efficiency to deploy an approximate Bayesian Computation sequential Monte Carlo (ABC-SMC) algorithm for fitting the in situ magnetic flux rope measurements. This ABC-SMC algorithm is a Bayesian inference algorithm that is not only capable of generating global best-fit solutions but also estimates constraints on the model parameters. The algorithm can be additionally fine-tuned by incorporating priors from either auxiliary measurements or physical considerations. The specific implementation of the fitting algorithm is easily extendable to multi-point constellations which will facilitate the study of multi-point events.

Due to the novelty of this algorithm in the context of studying CMEs, \edit1{we will briefly show a numerical test, using synthetically generated measurements, to show that our method delivers the correct results in the best-case scenario when the ground truth is known}. \edit1{Additionally we apply our model and algorithm to a flux rope observed in situ by PSP} on 2018 November 12 at 0.25 AU shortly after its first close fly-by of the Sun to show that our method is also applicable to real world scenarios.

This study is structured as follows. In Section \ref{sec:model} we introduce our improved 3DCORE model and provide an overview of the basic assumptions that we use. Section \ref{sec:methods} describes the motivation for the usage of the ABC-SMC algorithm, gives a very brief introduction to the topic and goes into detail on our specific implementation. Section \ref{sec:methods_synthetic} shows the results of our ABC-SMC algorithm with respect to synthetic measurements. \edit1{In Section \ref{sec:results} we apply the 3DCORE model and the ABC-SMC algorithm to the flux rope observed by PSP and interpret the results}. We discuss the applications of our novel approach in Section \ref{sec:conclusion}.

\section{Model} \label{sec:model}
In this section we give an in-depth description of the 3DCORE model and the accompanying data-generating process. We repeat most of the basic concepts of the model, which can also be found in the original paper by \citet{Moestl_2018}, for the purpose of clarity. The model serves to empirically describe and reproduce the general properties of CME flux rope measurements as we believe them to be. We attach ourselves to the concept of a bent flux tube that is connected on both sides to the Sun, describing the flux rope on a global scale. The self-consistency of this global picture can be tested by applying it to multi-point events, i.e. flux ropes that were measured by multiple spacecraft at significantly different positions. Such a study is well beyond the scope of the current paper, but the methods that we introduce can very easily be extended to multi-point analysis and we therefore lay the necessary groundwork for the future.

The solar wind environment heavily influences the evolution of any CME during its propagation within the interplanetary medium. As the solar wind is far from uniform in density or speed, specifically in the angular component, it is expected that the global shape of the CME is continuously deformed. Due to kinematic interactions, the flux rope is also expected to be flattened in the direction of propagation \citep{Riley_2004}. For our purposes we limit ourselves to describing the solar wind background using a single global speed value and associated global density coefficient. We assume that the geometry of our coronal mass ejection expands self-similarly during its slowed down or accelerated propagation without any deformation effects. The kinematical flattening effect is approximated using a constant elliptical cross-section.

The magnetic field is inserted into our flux rope shape using the same procedure as for local analytical magnetic field models. The specific implementation allows us to insert any arbitrary analytical magnetic field model into our chosen shape with only some minor modifications. This allows us greater freedom in choosing the magnetic field models for analysis, and future work will include extensive comparisons between different models.

In summary, we split the 3DCORE model into three different components. These are namely the shape model that describes the global geometry, the propagation model that describes the interaction with the background solar wind and the self-similar expansion, and the magnetic field model that inserts an analytical magnetic field into our chosen shape.

\subsection{Shape Model} \label{sec:model_shape}
\begin{figure}[h]
\includegraphics[trim=240mm 225mm 200mm 200mm, clip, width=\linewidth]{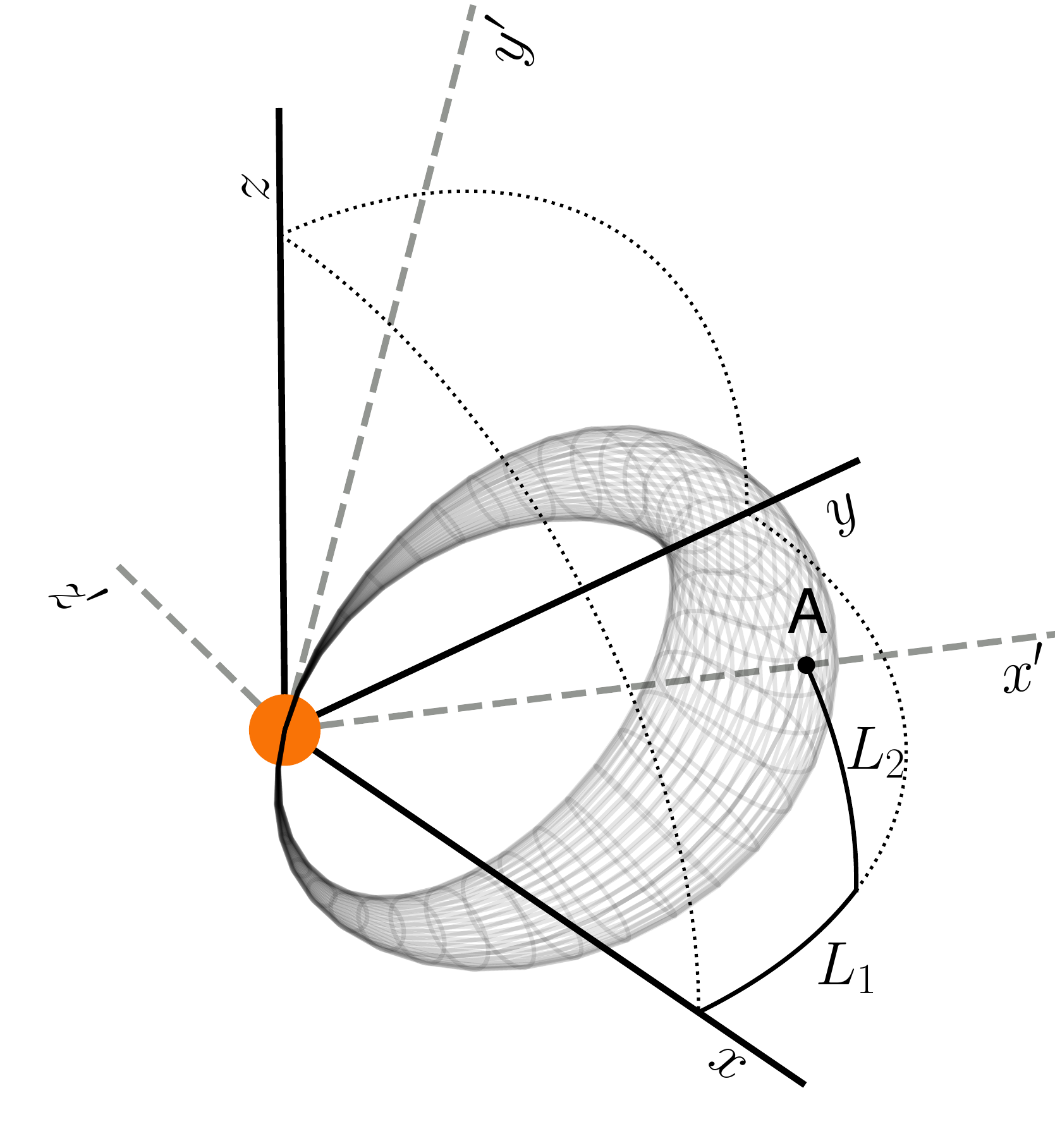}
\caption{\label{fig:11_shape} Torus-like geometry of the 3DCORE shape model with both cartesian coordinate systems. Additionally shown is the apex point A and the longitude $L_1$ and latitude $L_2$. The Sun (shown in orange) sits at the origin of the coordinate systems.}
\end{figure}

The global shape of our flux rope is described using a custom curvilinear coordinate system, further denoted as $Q$, that is defined using the parametrization shown in Eqs. \eqref{eq:model_shape_f}. 
\begin{subequations}
\label{eq:model_shape_f}
\begin{align}
x' &= -\left[\rho_0 + r\rho_1 \left(\sin{\frac{\psi}{2}}\right)^2\sin{\phi}\right]\cos{\psi} + \rho_0\\
y' &=\left[\rho_0 + r \rho_1\left(\sin{\frac{\psi}{2}}\right)^2\sin{\phi}\right]\sin{\psi}\\
z' &= \delta r \rho_1 \left(\sin{\frac{\psi}{2}}\right)^2 \cos{\phi}
\end{align}
\end{subequations}
The mappings $f': Q \mapsto \mathbb{R}^3$ and $g': \mathbb{R}^3 \mapsto Q$ define the forward and backward coordinate transformations that are required to transform $Q$-coordinates into Cartesian coordinates and vice-versa. The coordinates in $Q$ are denoted as $r \in [0, \infty),\ \psi \in [0, 2\pi),\ \phi \in [0, 2\pi)$. The inverse mapping $g'$ is shown in Eqs. \eqref{eq:model_shape_g}.
\begin{subequations}
\label{eq:model_shape_g}
\begin{align}
\tan{\psi} &= - \frac{y'}{\rho_0 - x'},\\
\tan{\phi} &= \frac{z'}{\delta \left(\sqrt{(\rho_0 - x')^2 + y'^2} - \rho_0\right)},\\
r &= \frac{\sqrt{(\rho_0 - x')^2 + y'^2} - \rho_0}{\left(\sin{\frac{\psi}{2}}\right)^2\rho_0 \cos{\phi}}.
\end{align}
\end{subequations}
This curvilinear coordinate system is based on the toroidal coordinate system. The parameters $\rho_0$ and $\rho_1$ define the major and minor radius of the base torus. The $\delta$ parameter controls the ellipticity of the cross-section. A value of $\delta > 1$ corresponds to a ``flattened'' cross-section as one would expect from flux ropes due to kinematic effects (some other papers use an inverse definition). The flux rope structure itself is defined by the implicit volume $r \leq 1$. This leads to a torus-like shape, with an elliptical cross-section of variable size that is permanently attached to the point of origin which acts as the Sun on the thinnest side. The resulting geometry is illustrated in Figure \ref{fig:11_shape}.

The parametrization is chosen so that the frontal part of the flux rope geometry, which indicates the direction of propagation, always points towards the positive $x'$-axis. In order to allow for general propagation in any direction we introduce a rotation $R$ that is defined by three further parameters. These are the latitude $L_1$, the longitude $L_2$ and the inclination, tilt or orientation $O$. They represent the direction of the propagation of the flux rope structure and its respective orientation. By our definition, an inclination value of $0^\circ$ (and $180^\circ$) lies within the XY plane of the rotated Cartesian coordinate system. The final mappings that we use in order to transform from general Cartesian coordinates, which are ideally coordinates in an inertial reference frame, are defined by: $f = R \circ f',\ g = g' \circ R^{-1}$. Unless otherwise stated we will be using the Heliocentric Inertial (HCI) coordinate system as default.

The fact that $g'$ can still be given in an analytical form for this particular shape is of great advantage. It allows extremely simple collision detection and evaluation of the magnetic field when given a magnetic field function $B(r, \psi, \phi)$ in Q-coordinates. This approach leads to a significant computational speed-up and is the most important improvement when compared to the implementation used in the original 3DCORE paper \citep{Moestl_2018}, as it opens up the possibility of using more computationally expensive methods for analysis. In the general case, when using an arbitrary parametrization for the shape, this is no longer necessarily true and numerical approximation schemes must be used.

\subsection{Propagation Model}
\label{sec:model_propagation}
Flux rope propagation is implemented  by adding time dependence to the shape parameters $\rho_0$ and $\rho_1$. This time dependence takes the following form:
\begin{eqnarray}
\rho_1(t) &=& \frac{D_\textrm{1AU} R_{\textrm{apex}}^{1.14}(t)}{2},\\
\rho_0(t) &=& \frac{R_{\textrm{apex}}(t) - \rho_1(t)}{2},
\end{eqnarray}
where $R_{\textrm{apex}}(t)$ is the distance of the apex point \edit1{A} to the Sun and  $D_\textrm{1AU}$ is the diameter of the flux rope cross-section at the widest point and at 1~AU.
For the propagation and interaction with the solar background wind, we only consider the kinematics of the apex point. The rest of the flux rope structure expands with the apex point in order to preserve the condition of self-similarity over time. The distance of our apex point is described using a drag-based model, based on \cite{vrsnak_2013}, with the following analytical form:
\begin{eqnarray}
\label{eq:model_propagation_apex}
R_{\textrm{apex}}(t) = &\pm& \frac{1}{\gamma} \ln\left[({1 \pm \gamma (V_0 - V_w)t}\right]\nonumber\\ &+& V_w t + R_0
\end{eqnarray}
where $V_0$ is the initial CME velocity, $V_w$ is the background solar wind speed, $R_0$ is the initial apex distance from the sun and $\gamma$ is the solar background wind drag-coefficient. This propagation model is unchanged from \citet{Moestl_2018}.
An important aspect of this particular drag-based model is the fact that there is an analytical expression for $R_{\textrm{apex}}(t)$. We can therefore directly evaluate $\rho_0(t)$ and $\rho_1(t)$ at any point in time without simulating the propagation from the start. The computational time needed for a single 3DCORE simulation therefore scales linearly with respect to the number of chosen time points at which the magnetic field is to be simulated.

\subsection{Magnetic Field Model \& Embedding}
Extensive work has been performed finding analytical magnetic field solutions for various flux rope geometries. The most common cases are fully analytical or approximate solutions for cylindrical \citep{Lundquist_1950, Gold_1960} or toroidal geometries \citep{Hidalgo_2012, Vandas_2017}. Additional approximate solutions exist for elliptical-cylindrical geometries \citep{Hidalgo_2002, Nieves_Chinchilla_2018}.
Since no solutions exist for our model shape we are forced to adapt an existing solution to our needs. In this particular case we choose to embed the toroidal  Gold-Hoyle-like solution described in \cite{Vandas_2017}. In normal toroidal coordinates this magnetic field takes the following form:
\begin{subequations}
\label{eq:model_magnetic_vandas}
\begin{align}
B_r &= 0,\\
B_\psi &= \frac{B_0}{1 + b^2 r^2},\\
B_\phi &= \frac{B_0 b r}{(1 + b^2 r^2)(1 + r \frac{\rho_1}{\rho_0} \cos \phi)}.
\end{align}
\end{subequations}
where $b$ is the twist parameter. We can relate the twist parameter with the total number of twists in the structure using:
\begin{equation}
b = \tau \frac{\rho_1}{2\pi\rho_0}
\end{equation}
where $\tau$ is now the total number of twists along the entire torus. In the case of our slightly different shape we correct the twist factor using:
\begin{equation}
b = \tau \frac{\rho_1}{2\pi\rho_0} E(\delta) \sin\left(\frac{\psi}{2}\right)^2
\end{equation}
where $E(\delta)$ is \edit1{the circumference of an ellipse with an aspect-ratio of $\delta$ and a minor axis length of one}. As this circumference cannot be calculated analytically we use a numerical approximation.

\begin{figure}[h]
\includegraphics[trim=150px 125px 100px 175px, clip, width=\linewidth]{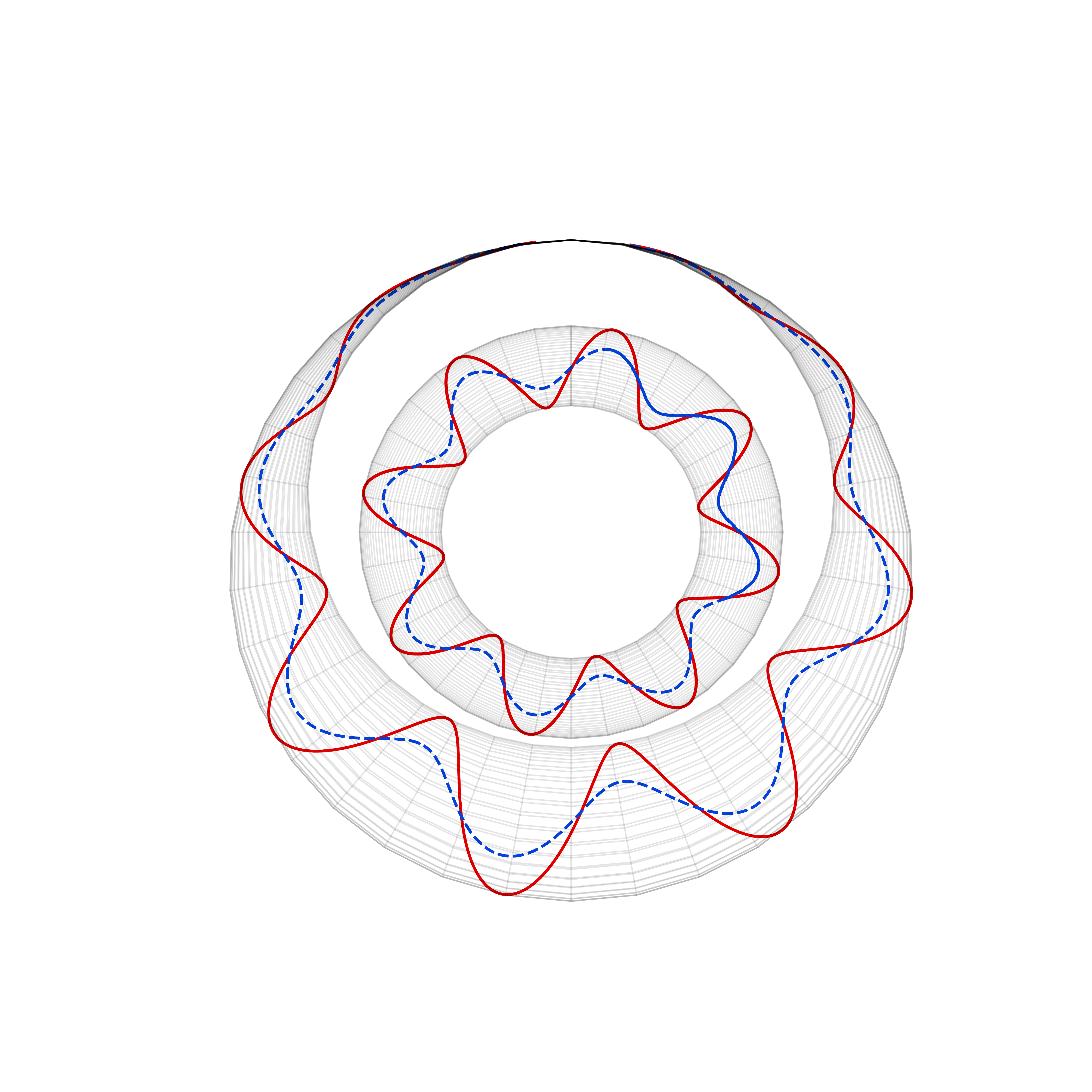}
\caption{\label{fig:13_model_field}Comparison between a normal toroidal geometry (inner) and our chosen global flux rope shape (outer). Within each structure we integrated along two magnetic field lines at different radial distances. The solid red magnetic field line is at a distance of $r=1$ (solid, red) and therefore part of the flux rope boundary. The dashed blue field line is at a central distance of $r=0.5$ (dashed, blue). For both cases we chose $\tau=8$ which results in a total of 8 twists over the entire structure.}
\end{figure}

\begin{figure}[h]
\includegraphics[trim=45px 0px 45px 50px, clip, width=\linewidth]{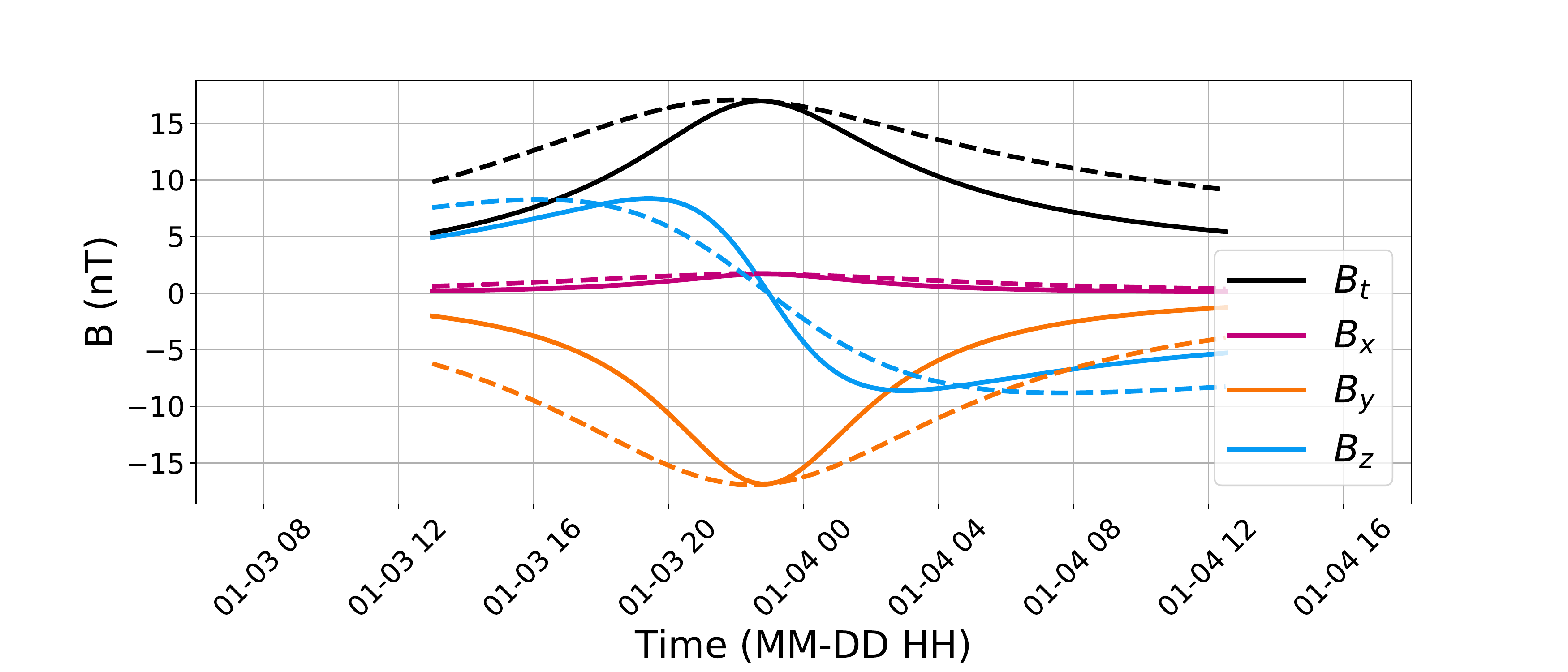}
\caption{\label{fig:13_model_profile} Two magnetic field profiles generated by the 3DCORE model simulating Earth-directed CMEs with twist values $\tau=8$ (solid) and $\tau=4$ (dashed). In both cases we see a characteristic rotation of the magnetic field in the $B_z$ component. We furthermore see a forward shifted peak for the magnetic field strength due to the continuous expansion of the CME structure.}
\end{figure}

\textbf{Figure \ref{fig:13_model_field}} shows traced magnetic field lines using our magnetic field model for both a normal torus and our torus-like shape. For each shape we trace two magnetic field lines positioned at coordinates $r=1$ and $r=0.5$ across the entire structure. This can be used as a visual verification for our magnetic field model as we used integer twist values for $\tau$. \textbf{Figure \ref{fig:13_model_profile}} shows two in situ magnetic profiles generated by simulated Earth-directed CMEs. In both profiles we see the characteristic rotation of the magnetic field in the $B_z$ component, and evolution of the total magnetic field strength with a peak that is slightly off center due to expansion.

\subsection{Noise}
\label{sec:model_noise}

An additional important aspect is the effect of noise or randomness. So far, most analytical flux rope models are deterministic models in the sense that they generate the same result given fixed initial conditions and parameters. Observed measurements, even if exactly modeled, will suffer from measurement noise or statistical fluctuations from the underlying physical process. This will limit the accuracy of any fitting procedure and may be the source of a significant amount of error or uncertainty. It may also be the case that different model parameter combinations generate very similar outputs. Adding any noise to two similarly generated measurements may then result in them being indistinguishable.

While reducing any inherent uncertainties is generally not possible, an estimation of the uncertainty would allow for the assessment of the quality of the model fit. As we will explain in detail in Section \ref{sec:methods}, this requires a model of the noise that appears in the measurements itself in the statistical sense. We make the highly simplified assumption of additive Gaussian noise, determined by the standard deviation $\sigma$, on the magnetic field measurements. This particular choice should be seen as more of a proof of concept than an accurate description of the underlying uncertainty that is present in the measurements. A more accurate description can only be obtained with an in-depth discussion on the specifics of the instruments and the relevant physics.

\section{Methods} \label{sec:methods}

In this section we introduce and discuss the numerical approach that we use to fit the in situ magnetic field measurements using the 3DCORE model that we described in the previous section. In particular we state the motivation and then introduce the ABC-SMC algorithm that we use for our subsequent analysis. In Section \ref{sec:methods_synthetic} we test the algorithm on a synthetic data set, generated by the 3DCORE itself, in order to verify that our implementation can correctly extract a known ground truth. This represents an analysis in the best-case scenario in which the model is fully capable of reproducing and describing the measurements. We could furthermore test how our results vary in the synthetic case for different levels of artificial noise and which model parameters can be inferred more accurately.

The most common approach to fitting MFRs with an analytical magnetic field model, cylindrical or toroidal, is to minimize a customly defined error metric using an iterative gradient-descent based minimization algorithm. A very popular error metric is the root mean square error that is either based on only the three magnetic field components or the components and the total magnetic field strength. This type of approach can be seen in the studies by \cite{Lepping_1990, Vandas_2017, Nieves_Chinchilla_2019} or papers that focus on comparing the various different methods \citep{Riley_2004_2}. These algorithms generally only guarantee convergence towards a local minimum and, while this approach has been shown to work very well for local magnetic field models, they tend to fail for global flux rope models \citep[e.g.][]{Isavnin_2016} due to the increased geometrical complexity. In the general case Monte-Carlo based methods are required that are capable of searching the full parameter space and finding global minima. The primary draw-back is a significant increase in the computational cost of the fitting algorithm itself.

Flux rope fitting methods that rely on the minimization of an error metric also only derive a single parameter estimate. The estimation of an error on the model parameters themselves requires multiple independent measurements of the same event (i.e. multi-point events with many satellites). Since the majority of events are only observed by a single spacecraft, in some rare cases by up to two or three, it is hard to perform any statistical analysis for these types of measurements. This error estimate is important in order to indicate the level of confidence in the derived results. Due to the complex 3D structure of the flux ropes that are only measured at one single point, i.e. strongly projected, there may be significant ambiguities and there is a strong possibility that a large range of different model parameters can reproduce essentially the same result. This ambiguity can be further amplified when attempting to analyze very noisy or strongly distorted flux ropes. 

These issues serve as a motivation to explore a different class of inference algorithms with the intent of at least partially alleviating the aforementioned issues. In this paper we will showcase a particular implementation on a Monte-Carlo based Bayesian inference algorithm that is capable searching the full valid parameter space (i.e. finding global minima) and generating error estimates on the resulting model parameters in the form of probability distributions.

\subsection{Bayesian Inference}

For our purposes we reformulate the fitting problem using Bayes' Theorem:
\begin{equation}
\label{eq:bayes}
p(\theta|y) = \frac{\mathcal{L}(y|\theta) p(\theta)}{p(y)}.
\end{equation}
where $\theta$ are the model parameters and $y$ is the data set that we use for the fitting procedure. The goal is to compute the posterior $p(\theta|y)$, which is a multi-dimensional probability distribution over the parameter space and gives the conditional probability of $\theta$ when given the data set $y$. It is important to keep in mind that the probability distributions encode our belief in which value a parameter takes and that the true underlying value is still fixed.

In order to compute the posterior we first need to compute the \textit{likelihood} $\mathcal{L}(y|\theta)$ and define a \textit{prior} $p(\theta)$. The $p(y)$ term can be safely ignored for all application as it simply reduces to a normalization factor due to $\int p(\theta|y) = 1$.

The prior $p(\theta)$ encompasses all information that we know about our model parameters before running any analysis. This can include constraints due to physical considerations or statistics from past events (generated from extensive CME catalogs). In the case where we do not want to include any additional information we are required to use non-informative priors which can be constructed in various different ways. Throughout the remaining work we will always use non-informative priors with uniform distributions over a certain range of interest. As we will briefly comment on later, this may introduce significant bias and error.

The second, and more important component, is the expression for the likelihood $\mathcal{L}(y|\theta)$. The likelihood gives the probability of generating $y$ given model parameters $\theta$. This highlights the need for a stochastic forward simulation model since $\mathcal{L}$ would simply reduce to a $\delta$ function in the case of a deterministic model. As we only inserted randomness into our model using random Gaussian noise (see Section \ref{sec:model_noise}) we are able to write down an analytical expression for $\mathcal{L}$. For a $k$-dimensional magnetic field measurement $y$, a simulation output $x = M(\theta)$ and a noise level of $\sigma$ the likelihood takes the following form:
\begin{equation}
\label{eq:likelihood}
\mathcal{L}(y|\theta) = \mathcal{N}(y - x, \sigma \mathbf{I}_k)
\end{equation}
where $\mathcal{N}$ is the k-dimensional multivariate normal distribution. In the more general case, and for more complicated models, it is not possible or highly impractical to find an expression for the likelihood.

\edit1{The posterior is then finally computed, or more correctly approximated, by sampling from the product $\mathcal{L}(y|\theta) p(\theta)$. Various different sampling algorithms have been developed to achieve this, with the most popular class of algorithms using Monte-Carlo Markov Chains (MCMC) \citep[e.g.][]{Hastings_1970}. While an algorithm like MCMC should in theory be applicable in our case, as we are given an analytical expression for the likelihood, we found it to not work well in practice. This is most likely due to our model being capable of generating non-results, i.e. CME's that completely miss the observer in space or time. These simulation runs do not allow the evaluation of a likelihood and therefore define invalid regions in the parameter space. These invalid regions require special handling and significantly slowed down our MCMC implementations to the point that it was not able to generate satisfactory results.}

\edit1{An alternative to MCMC are sequential Monte-Carlo (SMC) sampling algorithms \citep{Moral2006} which are also sometimes referred to as particle filters. Their primary advantage is that these algorithms are generally easily parallelizable, which is not the case for MCMC, leading to a significant speed-up for the sampling process. The SMC sampler can also better handle the invalid regions of the parameter space without some of the major pitfalls that may hamper the convergence of an MCMC, albeit still at a significant computational cost.  Despite these advantages we were still not able to generate satisfactory results using an SMC sampler as it, among other things, converged too slowly. While it should be expected that a more sophisticated SMC algorithm can overcome these issues the method would still be limited by the requirement for an analytical expression for the likelihood which may restrict the applicability for future models.
}

\edit1{For this reason we decided to make use of a more general class of algorithms known as approximate Bayesian Computation (ABC), sometimes also referred to as likelihood-free algorithms. In ABC the likelihood is replaced by a summary statistic. They therefore do not require an expression of the likelihood, which is their primary advantage, and as such greatly expands the realm of models to which they can be applied. Their only requirement is a more or less accurate and more importantly fast forward simulation that can reproduce the observed measurements.}

\subsection{Approximate Bayesian Computation}

For a more extensive introduction to ABC we refer the interested reader to \cite{Sisson_2018}. The principle idea behind any ABC algorithm is to bypass the likelihood $\mathcal{L}$ using a distance metric $\rho(x,y)$, which can be defined in various ways, that measures the difference between data $y$ and simulation outputs $x = M(\theta)$. The posterior is approximated using the ensemble:
\begin{equation}
\label{eq:abc_estimate}
p(\theta|y) \approx \lim_{\epsilon \rightarrow 0} \{\theta_i | \rho(x_i = M(\theta_i), y) < \epsilon\}
\end{equation}
where $\epsilon$ is a threshold value and $\{x_i\} = \{M(\theta_i) | \theta_i \sim p(\theta)\}$ \citep{Tavare_1997, Beaumont_2002}. The set of $\theta_i$ used in the ensemble in Eq.\ \eqref{eq:abc_estimate} is initially drawn from the prior $p(\theta)$ and only the parameter candidates $\theta_i$ that satisfy $\rho(x_i = M(\theta_i), y) < \epsilon$ are used to approximate the posterior $p(\theta|y)$. This is the simplest form of an ABC algorithm and is called the rejection algorithm. While it illustrates the core principles of any ABC algorithm it is of practically no use except for the most simplistic cases or toy models.

In practice the rejection algorithm suffers from multiple problems. For smaller $\epsilon$ values, or as $\epsilon$ tends towards zero, the number of rejected parameter candidates $\theta_i$ becomes extremely large. This generally leads to the existence of a threshold value $\epsilon_\text{min}$ below which no accepted candidates can be reliably found. Another source of error is the distance metric $\rho(x, y)$. Using extra-large or multi-dimensional data sets or model outputs for the $\rho$ statistic will additionally increase the rejection rate of the ABC algorithm due to the curse of dimensionality. This problem is commonly circumvented by using a summary statistic $S$ so that $\rho=\rho(S(x), S(y))$ where $S$ reduces the dimensionality and complexity of the data set and model outputs. While the usage of this summary statistic is highly beneficial for computational efficiency \edit1{it can be} another additional source of bias error \citep[e.g.][]{Prangle_2015}. Lastly, sampling from the entire prior $p(\theta)$ is highly inefficient as large regions of the prior space can be identified as being of very little interest when using wide priors.

While some of the inherent deficiencies of the basic ABC algorithm are hard to tackle, one can significantly improve the sampling process by which candidates $\theta_i$ are drawn from $p(\theta)$. Various more sophisticated ABC algorithms have been proposed such as those based on Markov chain Monte Carlo chains \citep[MCMC,][]{Marjoram_2003} or particle filters \citep[SMC,][]{Sisson_2007, Beaumont_2008} (sequential Monte Carlo). For our analysis we opted to use an ABC sequential Monte Carlo (SMC) algorithm for which we will explain the implementation in detail in the next section.

\subsection{ABC-SMC Implementation}

The  first choice that we make for our implementation is the definition of the distance metric $\rho$ and the summary static $S$. For our study we opted to simply use the root mean square error between the simulation result $x$ and reference data $y$ at K time points $t_i$:
\begin{equation}
\rho_\textrm{RMS}(x, y) = \sqrt{\frac{1}{K} \sum^K_{i=1}(x_i - y_i)^2 }
\end{equation}
where $x_i$ and $y_i$ are magnetic field vectors. The summary static $S$ reduces the full time-series $y$ to only a handful of measurements at significant time intervals $\{y_0, \hdots, y_{K-1}\}$. We found $K$ values of around a dozen to be sufficient for our analysis and this generally corresponds to time intervals of one to a few hours if the time points are uniformly spaced out across the entire flux rope. Due to the setup of our model, as was detailed in Section \ref{sec:model_propagation}, we can directly generate a simulation result at any time point. Using smaller $K$ values is therefore significantly faster than using larger ones with a linear relationship for the computational complexity.

This statistic has a significant issue when two signatures with a time-shift are compared. In order to remedy  this issue we further introduce two control points $t_S, t_E$ with $t_S$ being set just before the CME arrives $t_E$ being set after it has passed. These control points are generally inserted a few hours before and after the start and end of the flux rope. Any sample is rejected if the observer is within the synthetic flux rope at times $t_S$ or $t_E$. Additionally, we require that any sample generates a valid magnetic field measurement at the $K$ reference time points. This only allows flux rope measurements with a small time-shift and similar duration to be accepted. Assuming that the model returns a null vector when the observer is outside of the flux rope, the final summary statistic can be described as follows:
\begin{equation}
    \rho(x, y) = 
    \begin{cases}
    \phantom{\infty,} &x_{S} \neq 0\\
    \infty, &x_{E} \neq 0\\
    \phantom{\infty,} &\exists i: x_i = 0\\
    \rho_\textrm{RMS}(x, y), &\textrm{otherwise}
    \end{cases}
\end{equation}
which ensures that any simulations with signatures that vary greatly in duration compared to the reference measurement are rejected. The time discretization used in the summary statistics is most likely a significant source of error when performing inference, and some of the likely effects will be discussed later in the examples.

The basic idea of the ABC-SMC algorithm is to iteratively approximate the posterior using intermediate distributions with larger threshold values $\epsilon$. Furthermore each intermediate approximation of the posterior is generated by sampling parameter candidates from the previous approximation instead of the full initial prior. In the first iteration the ABC-SMC algorithm matches the simple rejection algorithm. We sample a set of candidate parameters $\{\theta_i^0\} \sim p(\theta)$ and generate the first intermediary distribution using $P^0 = \{\theta^0_i | \rho(M(\theta^0_i), y) < \epsilon_0\}$. The sampling is repeated until the set $P^0$ reaches a predetermined size $N$ that depends on the required resolution. Each ``particle'' within $P^0$ is then assigned the same weight $\omega_i^0= \frac{1}{N}$. Due to the specific construction of our distance metric $\rho$ the exact value that we chose for $\epsilon_0$ is not as important as most candidates will be rejected due to a significant time-shift when comparing the flux rope signatures. For our implementation we found $\epsilon_0 = \rho(0, y)$ to be an initially high enough starting value.

In each successive iteration we generate a new set of candidate parameters by sampling from the previous intermediary instead of the prior, i.e. $\{\theta_i^j\} \sim P^{j-1}$. As the intermediary $P^{j-1}$ is only given in approximate form by $\{\theta_i^{j-1}\}$ specific methods must be used to correctly draw new candidate parameters. This can be done by randomly picking (accounting for the weights $\omega_i^{j-1}$) a particle $\overline{\theta}_i^j$ from the intermediary $P^{j-1}$ and perturbing the selected particle via a perturbation kernel. Different variants of this method are explained in detail and compared in \citet{Filippi_2011}. The most common approach is to perturb $\overline{\theta}_i^j$ by a vector drawn from a multivariate normal distribution that can be constructed by computing the covariance matrix of the overall distribution $P^{j-1}$. Larger perturbation kernels are capable of probing larger areas of the parameter space and lessen the probability that the iterative algorithm finds itself stuck in a local minimum. On the other hand smaller kernels will lead to a higher probability of drawn candidates \edit1{being accepted} which leads to faster convergence (but not necessarily towards the correct result). 

\edit1{
For our case we opted to use a transition kernel based on M-nearest neighbours \citep{Filippi_2011} where the co-variance matrix is computed from only half of all particles within $P^{j-1}$ that are closest to $\overline{\theta}_i^j$ ($M=N/2$). This choice delivered significantly better results compared to the standard method of using the full co-variance matrix due to the particular shape of the intermediary distributions. Due to our model, there are various degeneracies that are expected to occur at large $\epsilon$ values that make the entire sampling process extremely inefficient when using a large transition kernel. The only downside with our choice is that special care needs to be taken that the algorithm converges correctly and does not get stuck in a local minimum due to insufficient exploration of the parameter space.
}

After generating a new intermediary distribution $P^j$ we assign weights to each particle according to \citet{Beaumont_2008}:
\begin{equation}
\omega_{i}^{j+1} = \frac{p(\theta_i^{j+1})}{\sum_{i'=0}^N \omega_{i'}^j K(\theta_{i}^j | \theta_{i'}^{j-1})}
\end{equation}
where $K(\cdot|\cdot)$ describes the transition probability given by the perturbation kernel.

The final component required for the algorithm is the determination of the threshold values for $\epsilon$. For this purpose we use an adaptive scheme based on quantiles \citep[e.g.][]{Lenormand_2011}. The $\epsilon_j$ value is computed as the $\alpha$-th percentile from the set $\{\rho(M(\theta_i^j), y)|\theta_i^j \in P^j\}$. We generally found values of $\alpha \in [0.2, 0.5]$ to work well.

Any of the generated intermediary distributions can serve as an approximation for the final posterior. As such we furthermore need a criterion when to abort the iterative algorithm. In practice, as the threshold value $\epsilon$ continues to decline, the algorithm becomes successively slower as most parameter candidates (and simulation results) are rejected. In our implementation we stop the algorithm when the number of iterations exceeds a predetermined amount or the accept-reject ratio (the number of drawn samples that are accepted compared to those that are rejected) falls below a predetermined value. Both of these two abort conditions can be modified as required. \edit1{For any reasonably complex situation ABC algorithms are not expected to fully converge or converge very slowly. As such the main criteria for choosing the stopping point are computational and time constraints.}

\subsection{ABC-SMC Synthetic Example} \label{sec:methods_synthetic}

We provide a detailed illustration of the type of results our ABC-SMC algorithm can deliver by applying the algorithm on a synthetic measurement. For this purpose we generate a magnetic field measurement using the 3DCORE model itself, with which we will then perform the analysis. This type of test is important in order to verify that our algorithm delivers correct results under ideal conditions.

\begin{figure}[h]
\includegraphics[trim=15mm 5mm 15mm 5mm, clip, width=\linewidth]{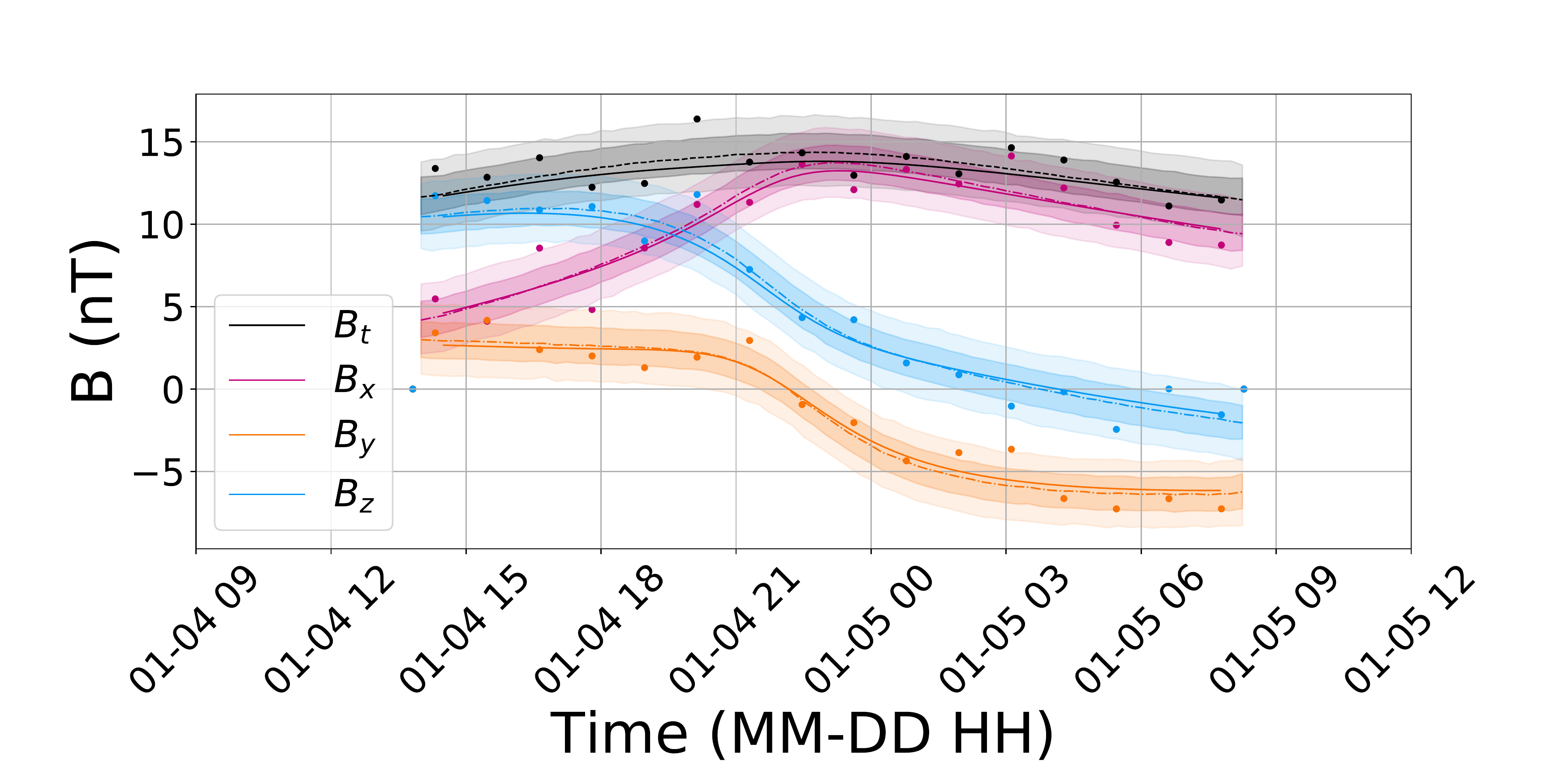}
\caption{\label{fig:30_synthetic_fitting} Synthetic flux rope measurement generated by the 3DCORE model using the model parameters from Table \ref{table:abc_fiducial_parameters}. The solid lines show the three magnetic field components generated by the model. The dotted points show the noisy measured magnetic field as it would be used by our ABC algorithm with 1 nT noise. The dashed line with the shaded area shows the resulting fit with the 1-$\sigma$ and 2-$\sigma$ spread that is generated by the ensemble representing the posterior.}
\end{figure}

\textbf{Figure \ref{fig:30_synthetic_fitting}} shows the synthetic flux rope that we used for this test and the resulting fit from our ABC-SMC algorithm. The smooth solid line shows the underlying signal from the 3DCORE model. \edit1{The dotted markers show the actual noisy measurement used for the analysis}. The resulting fit is the dashed line with a shaded area representing the 1-$\sigma$ and 2-$\sigma$ spread of the ensemble solution.

\begin{figure*}[ht]
\plotone{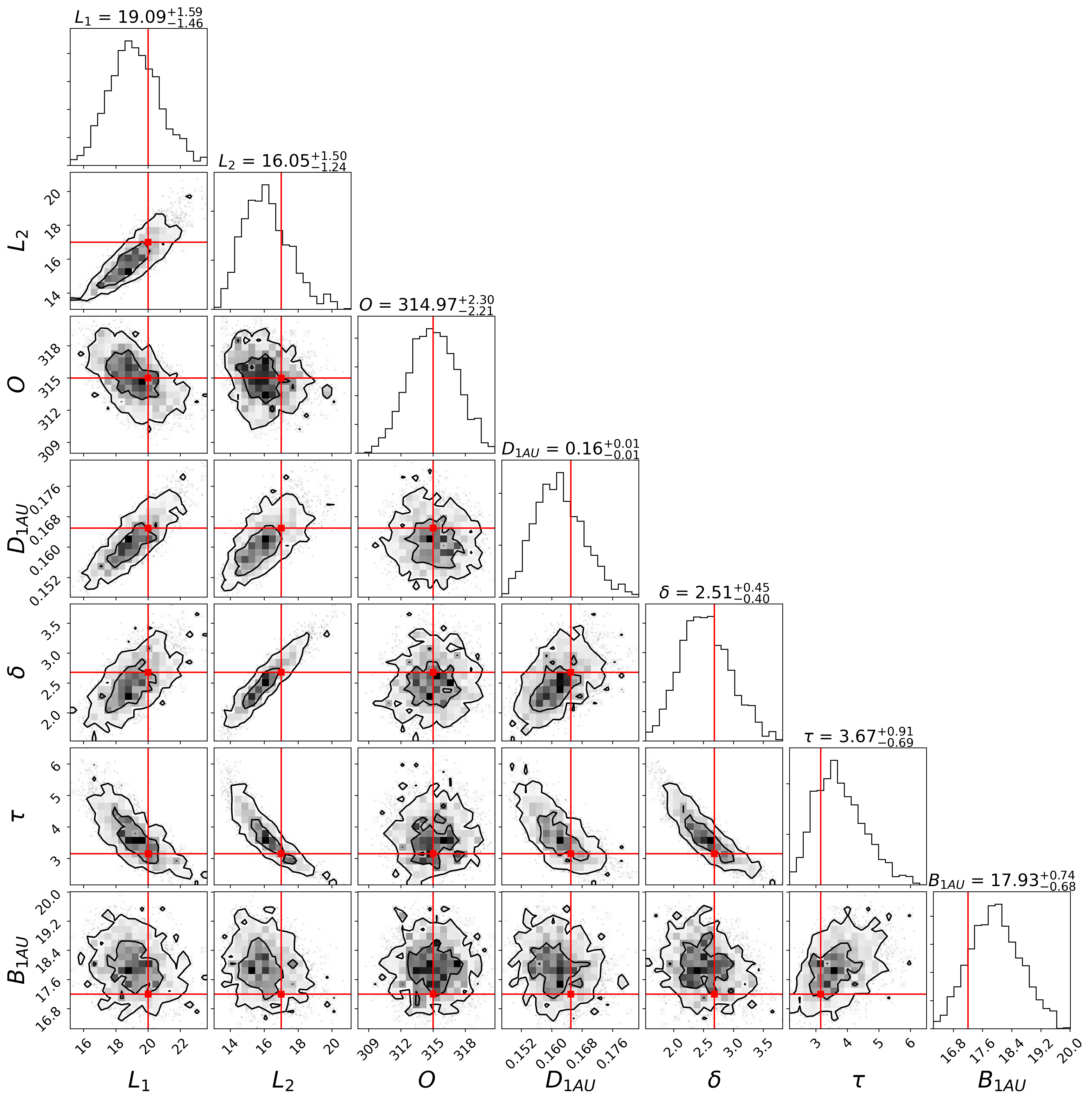}
\caption{\label{fig:30_synthetic_params} Scatter-plot matrix of the results of the ABC-SMC algorithm using the synthetically generated data set from Table \ref{table:abc_fiducial_parameters}. The diagonal shows the approximated posterior distributions for each free parameter. The other plots show the two dimensional posteriors for each parameter combination. The horizontal and vertical lines mark the underlying ground truth which was used to generate the synthetic data. Furthermore we marked the 1-$\sigma$ and 2-$\sigma$ boundaries using contours in the 2D posteriors. This result showcases the best-case scenario for our analysis as the model can fully describe the underlying measurement and can exactly determine the boundaries of the flux rope within our measurement. The uncertainties shown here are only generated by our chosen noise level and inherent uncertainties from the model.}
\end{figure*}

The synthetic measurement used was generated by the 3DCORE model using the model parameters shown in Table \ref{table:abc_fiducial_parameters}. The single parameter that is not listed in the Table is the initialization time $t_0$, which is set to 2018-01-01 00:00. In order to simplify the analysis for the algorithm we furthermore fixed multiple parameters in the inference. We set the parameters $t_0, R_0, V_0, V_{sw}, b_s$ and $\sigma$ to their true underlying values. With the exception of the $\sigma$ parameter these parameters can be estimated using coronagraph or heliospheric imagers for any real event. 

\begin{table}[h]
\center
\begin{tabular}{ c c c c c c }
$L_1$ & $L_2$ & $O$ & $D_\textrm{1AU}$ &
 $\delta$ & $R_0$\\
 \hline
 $20^\circ$ & $17^\circ$ & $315^\circ$ & $0.165\,\textrm{AU}$ & $2.68$ & $20\,R_\odot$
 \\
 $V_0$ & $\tau$ & $B_\textrm{1AU}$ & $V_{sw}$ & $\Gamma$ & $\sigma$\\
 \hline
 675\,km s$^{-1}$ & $3.15$ & $17.2\,\textrm{nT}$ & 368\,km s$^{-1}$ & 0.65 & 1 nT
\end{tabular}
\caption{\label{table:abc_fiducial_parameters}Fiducial initial parameters used in our ABC-SMC mock example. Further parameters, not shown in the table are the initialization distance at 20 solar radii and the initial launch time which is set to 2012-01-01T00:00:00 UTC.}
\end{table}

The fitting is performed on a 16 hour interval, with the individual fitting time points and the corresponding noisy measurements shown as dots in Figure \ref{fig:30_synthetic_fitting}, using a total of 16 fitting points. The $t_S$ and $t_E$ markers are set to exactly 30 minutes before/after the flux rope is measured. The initial threshold value is calculated as $\epsilon_0 = 12.3\,$nT which is reduced to the final value of $\epsilon_{16}=1.7\,$nT after 17 iterations. The $\alpha$ hyper parameter was set to $0.25$ and the number of particles $N$ per iteration was set to 4096.

\textbf{Figure \ref{fig:30_synthetic_params}} shows a so-called scatter plot matrix of the remaining free parameters. From this result we can see that our algorithm is able to accurately infer the correct range where the ground truth is located. The parameters which control the orientation $L_1, L_2$ and $O$ are inferred with an extremely unrealistic accuracy of only a few degrees. The two parameters with the lowest accuracy are $\delta$ and $\tau$. As can be seen in the $\delta-\tau$ 2D scatter plot these two parameters are not independent, i.e. there exists a degeneracy. This is most likely an artifact of our magnetic field model due to the twist being modified by the factor $E(\delta)$. This degeneracy will severely limit the accuracy with which we can infer the twist parameter in flux ropes. Given a fixed $\delta$ parameter, possibly determined from auxiliary measurements, the accuracy on $\tau$ could be considerably improved.

As the ABC-SMC algorithm probes the entire parameter space of our model, the resulting uni-model probability distributions for the parameters show that are no multiple local minima, at least within the achieved $\epsilon$ threshold of the summary statistic, for this particular synthetic event. While there is no absolute guarantee that this is actually the case, we were able to gather evidence for the synthetic case by running the ABC-SMC algorithm multiple times with different random seeds (that control how the random samples are drawn).

\section{Results} \label{sec:results}

In this section we apply the 3DCORE model and the associated ABC-SMC algorithm to an observed flux rope measurement. For this purpose we select an event that was captured by PSP on 2018 November 11-12 shortly after its first fly-by at the Sun at approximately 0.25~AU. This event represents the in situ observation of a CME magnetic flux rope at the smallest heliocentric distance in space history. A recent detailed study of this event was also presented in \cite{nieves_chinchilla_2020}.

\begin{figure}[h]
\plotone{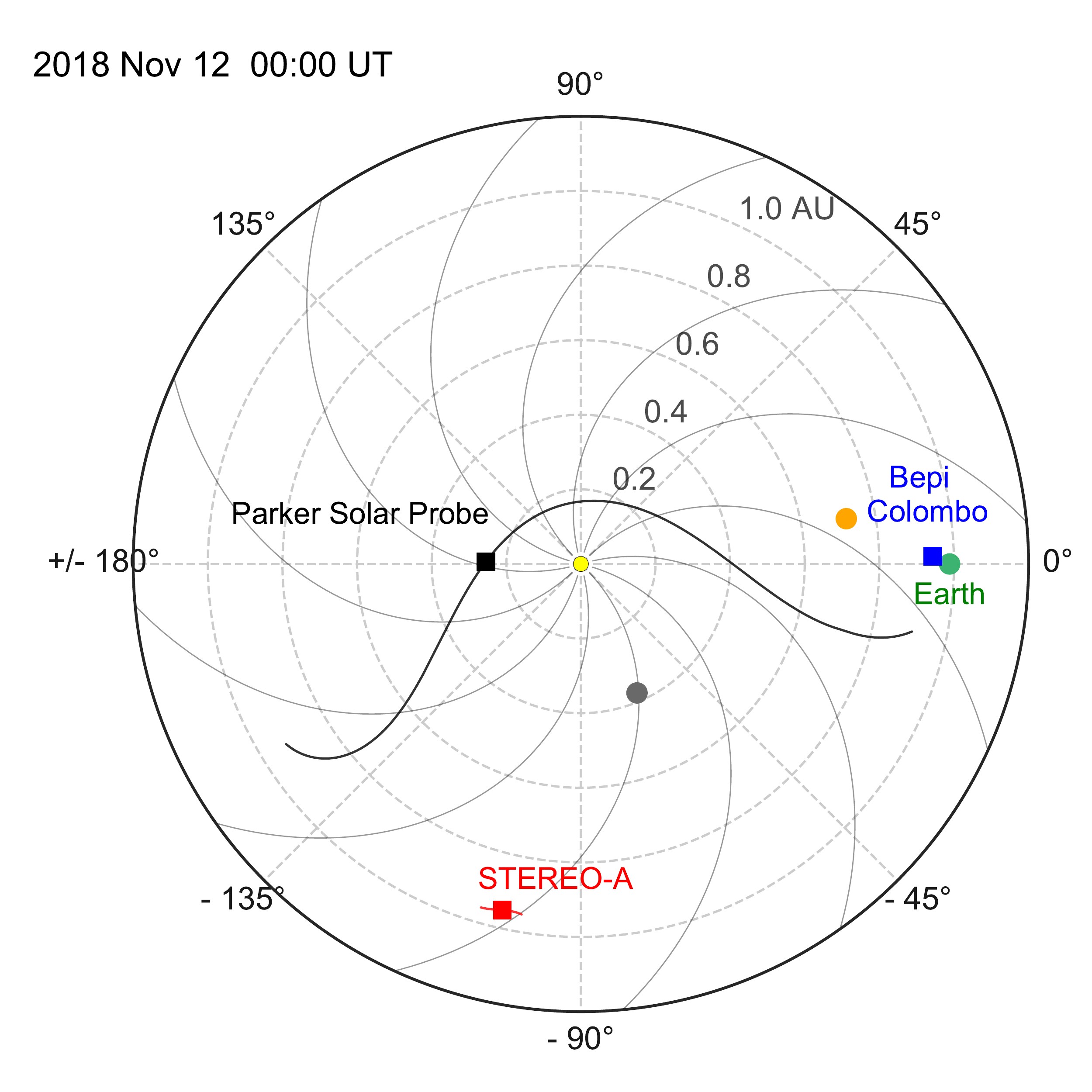}
\caption{\label{fig:40_positions}Overview of spacecraft and planet positions on 2018 November 12 00:00 UT. The positions of planets and spacecraft, indicated by the color code (Earth green, STEREO-A red, PSP black, and Bepi Colombo blue) are shown in an HEEQ coordinate system. The trajectories of PSP and STEREO-A are indicated as a solid line spanning from -60 to +60 days in time around the event, and the Parker spiral is plotted for a 400 km~s$^{-1}$ solar wind.}
\end{figure}

\textbf{Figure \ref{fig:40_positions}} shows the spacecraft and planetary constellation on 2018 November 12 00:00 UT. PSP was positioned almost exactly at the backside of the Sun as seen from Earth, at 178.6 degree heliospheric longitude, while STEREO-Ahead was close to quadrature, at -102.8 degree longitude from Earth, which is a favorable position for imaging CMEs that are either Earth directed or on the backside as seen from Earth.

\begin{figure}[h]
\plottwo{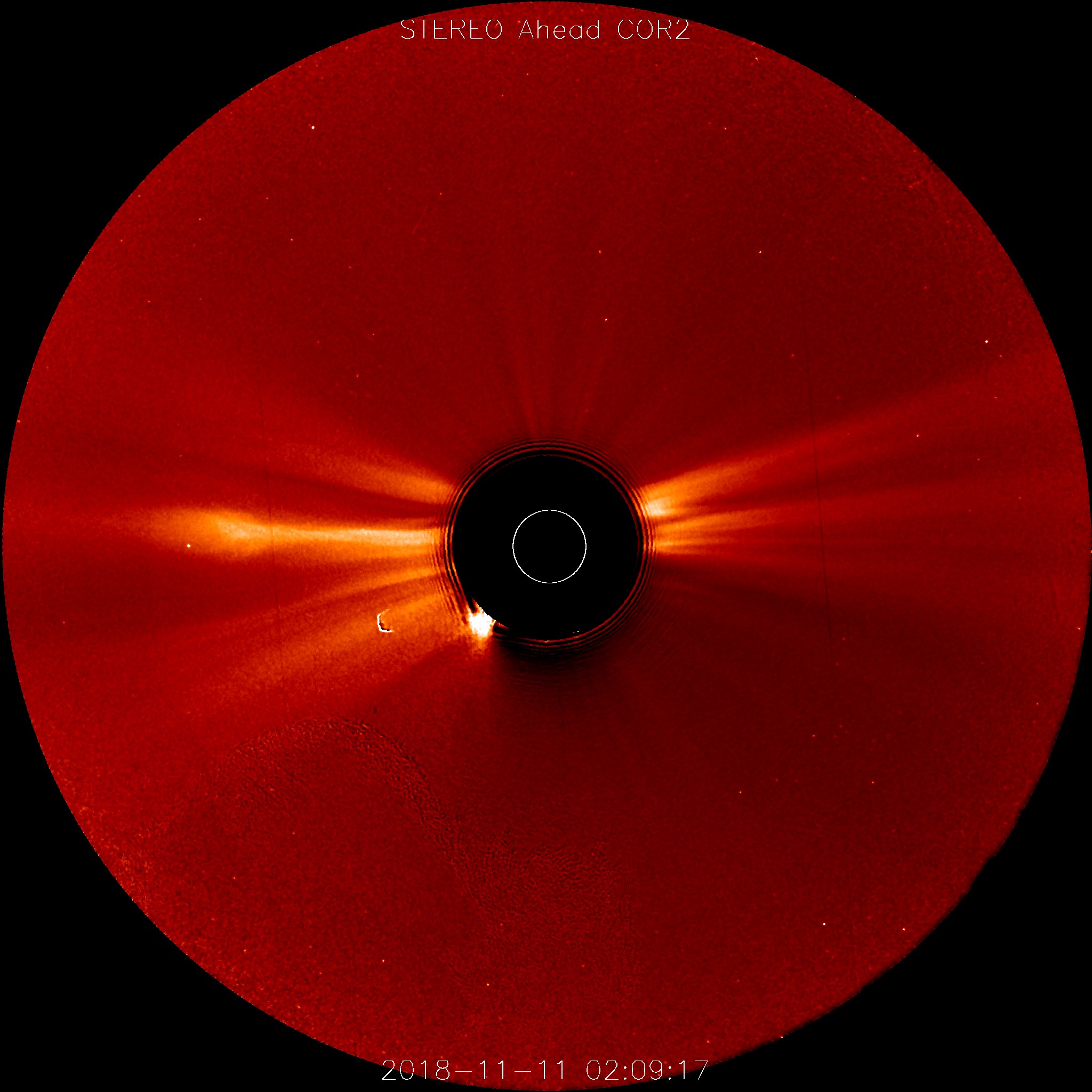}{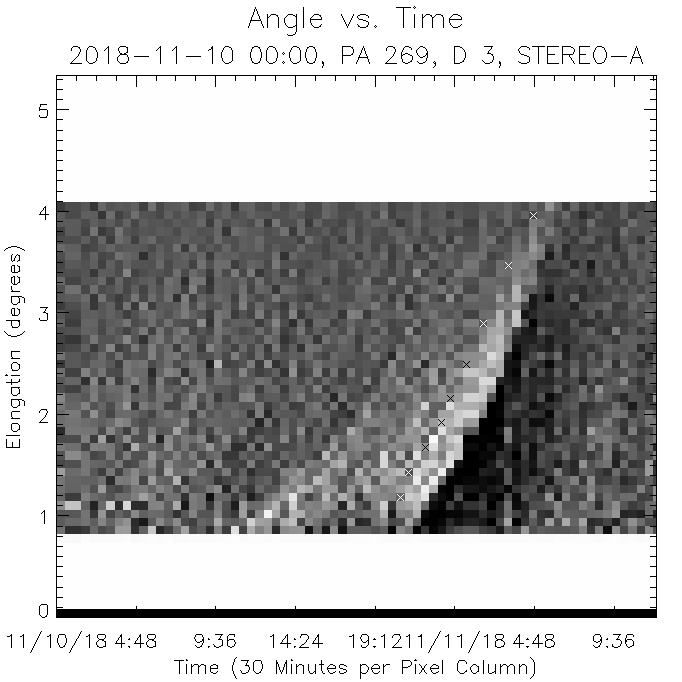}
\caption{\label{fig:40_satplot}On the left side, we show a representative image of the STEREO-Ahead COR2 coronagraph at 2018 November 12 02:09 UT that depicts a small CME structure propagating away from the Sun towards solar east. On the right side we show the generated Jplot image that shows the propagation of the CME front in detail over time. }
\end{figure}

\textbf{Figure \ref{fig:40_satplot}} shows an image from STEREO-Ahead's COR2 coronagraph at November 12 02:09 UT and the corresponding Jplot. The coronagraph shows a small CME structure propagating away from the Sun within the ecliptic at around 10 $R_\odot$ (left side on the image). Using the Jplot we estimated the CME velocity to be approximately $280 \pm 20$~km~s$^{-1}$ at 15~$R_\odot$.

\begin{figure*}[ht]
\includegraphics[trim=45px 25px 60px 60px, clip, width=\linewidth]{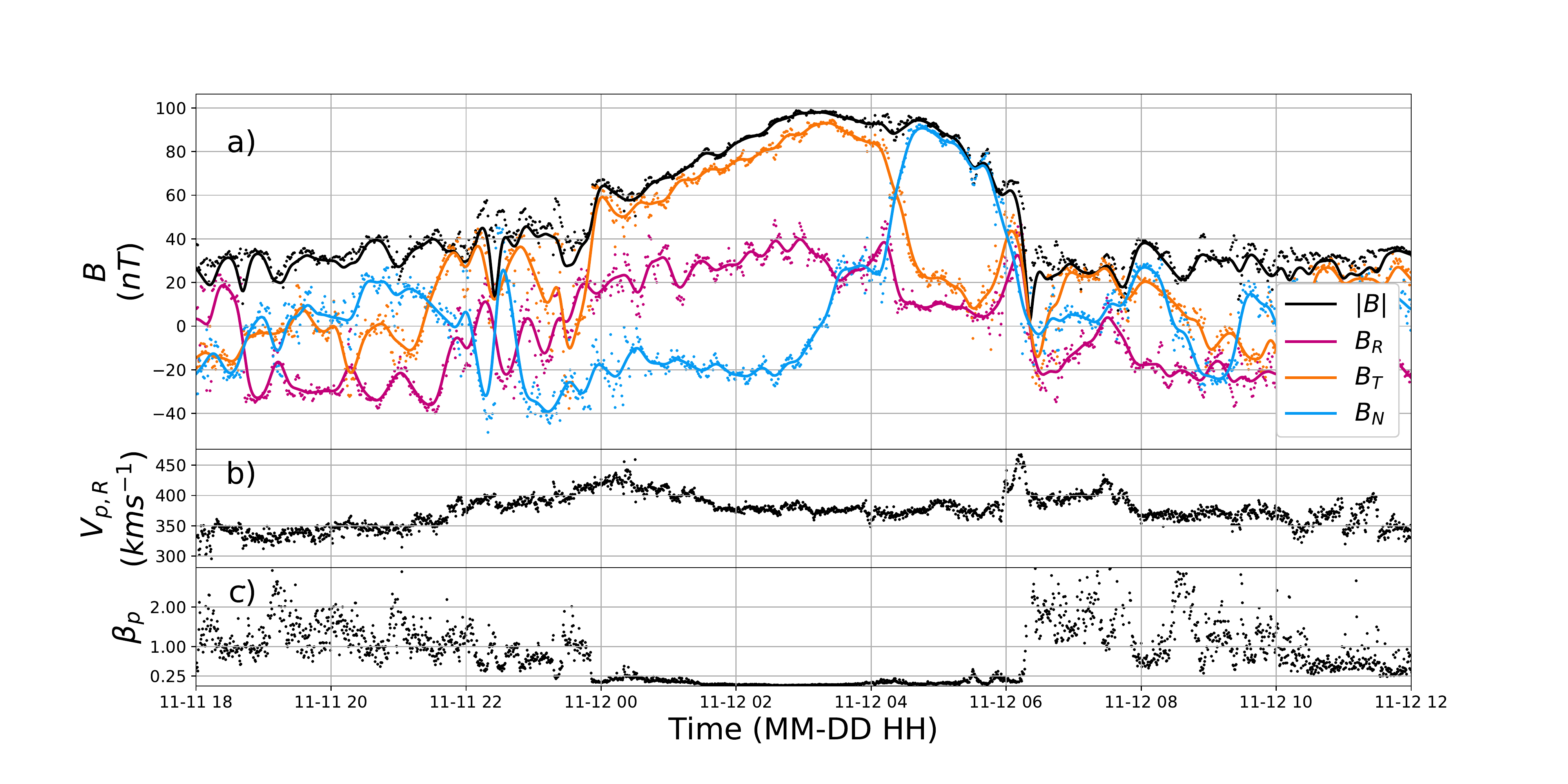}
\caption{\label{fig:40_event} Parker Solar Probe in situ measurements from 2018 November 11 18:00 UTC to November 12 12:00 UTC. The original L2 (magnetic field) and L3 (plasma) data are shown as dots. \textbf{a)} Magnetic field measurements from the FIELDS instrument in local RTN coordinates. For our analysis we use smoothed measurements which are shown with solid lines. The smoothing was performed using a Gaussian kernel with a 15-minute width. \textbf{b)} Radial proton velocity component measurement from the SWEAP instrument. These measurements show that the coronal mass ejection impacts PSP with up to $\approx$ 425 km s$^{-1}$. The $V_{p,T}$ and $V_{p,N}$ velocity components are negligible. \textbf{c)} Plasma $\beta_p$ coefficient we use as our primary indicator for the presence of the magnetic flux rope, from which we set the rope duration from 2020 Nov 11 23:52 to Nov 12 06:13 UTC.}
\end{figure*}

\textbf{Figure \ref{fig:40_event}} presents the in situ measurements for the magnetic field, the proton bulk velocity and the resulting $\beta_p$ coefficient for the event around 24 hours after the CME was observed by COR2 on STEREO-Ahead. The magnetic field was measured by the FIELDS instrument \citep{Bale_2016} and the proton bulk speed, density and temperate by SWEAP \citep{Kasper_2016}. For our analysis we used the publicly available L2 data-set for the magnetic field and the L3 moments for the proton measurements. In the case of the magnetic field, we further applied smoothing in the form of a Gaussian kernel of width $\sigma=900s$ (solid line).

The measurements show clear characteristics of a transient flux rope event, in the form of a significantly enhanced magnetic field and a very small $\beta_p$ coefficient. The magnetic cloud can be identified by inspecting the $\beta_p$ coefficient, which describes the ratio in between thermal and magnetic pressure. We use the extremely small $\beta_p$ value, lower than 0.25 for most of the period in between 23:52 UT and 6:13 UT, as the indicator for the magnetic cloud interval.
The coronal mass ejection itself is slow with an estimated speed of $425\,\textrm{km}\,\textrm{s}^{-1}$ at the front edge. During its propagation over PSP the speed decreases to an average of $387\,\textrm{km}\,\textrm{s}^{-1}$, only slightly faster as the solar background wind speed of approximately $350\,\textrm{km}\,\textrm{s}^{-1}$ in the hours preceding the event.

As is described in detail in \cite{nieves_chinchilla_2020}, there are indications that the local measurements show two interacting structures. This may explain the strong distortion that is present in the observed flux rope that only undergoes little change from 23:52 until around 04:00 at which point the magnetic field changes very rapidly. Due to the limited scope of our study and the applicability of our model we will assume that there is only one single structure.

We apply the ABC-SMC algorithm to the PSP magnetic field measurements using the time range 2018-11-12T01:00 -- 2018-11-12T06:00 with 11 equidistant fitting time points at half-hour intervals. We furthermore set the $t_S$ and $t_E$ markers to 2018-11-12T00:00 and 2018-11-12T07:30 respectively, allowing for flux rope solutions with up to 7.5 hours in duration. The CME initialization is set at 2018-11-11T06:00 at a distance 15 $R_\odot$ from the Sun with a fixed initial velocity of 280~km~s$^{-1}$. These are the only initial conditions that we use for our algorithm, and all other parameters, including the noise level, are described by flat priors within sensible ranges.

The hyper parameters for the algorithm were chosen similarly to the synthetic example in Section \ref{sec:methods_synthetic}. The particle count was doubled  to $N=8192$, as a higher number of free parameters requires a higher resolution. The adaptive threshold value $\alpha$ was set to an aggressive value of $0.25$. The initial threshold value is computed to be around $\epsilon_0 = 80\,$nT.

\begin{figure*}[ht]
\includegraphics[trim=35px 0px 55px 5px, clip, width=\linewidth]{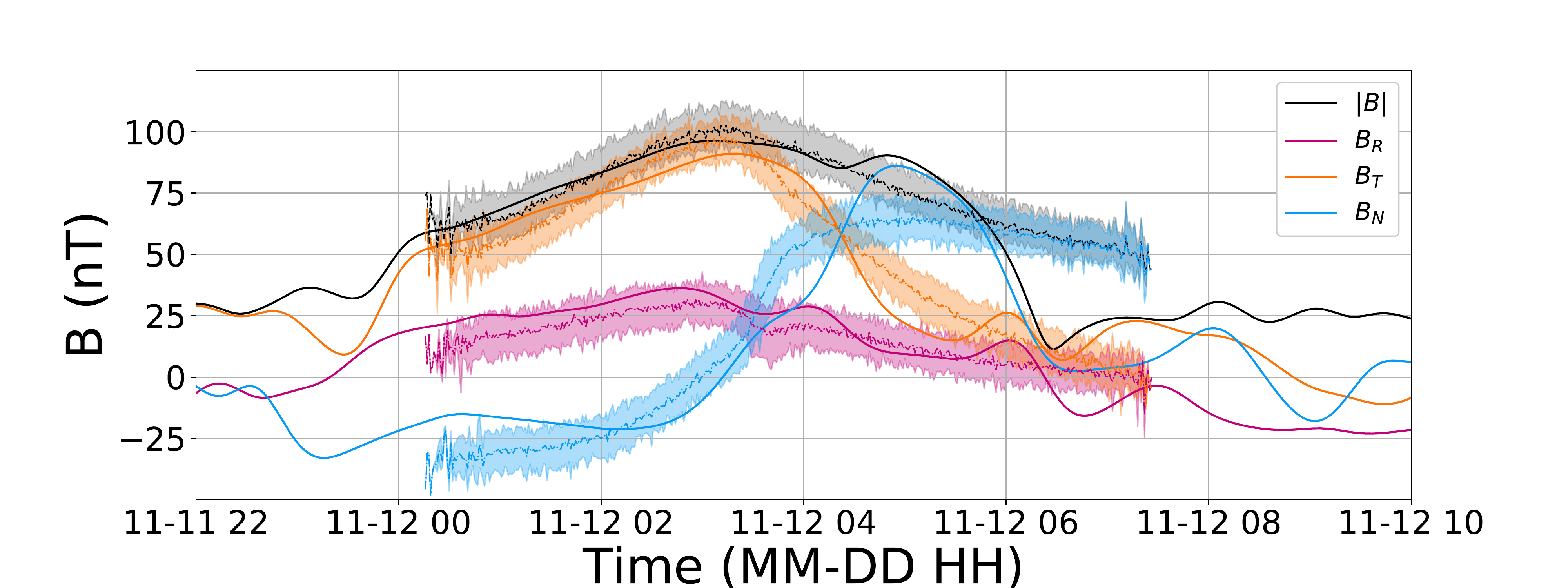}
\caption{\label{fig:40_fitting} PSP magnetic field measurements (solid) and the corresponding flux rope ensemble generated by the ABC-SMC algorithm (dashed). The shaded areas correspond to the 2-$\sigma$ spread of the ensemble. All components including the total magnetic field are well reproduced up to 04:00 UTC at which point the components, especially $B_N$, start to diverge. All ensemble members of the fit have a flux rope signature that lasts until 07:30 UTC which is the defined $t_E$ marker.}
\end{figure*}

\begin{figure*}[h]
\plotone{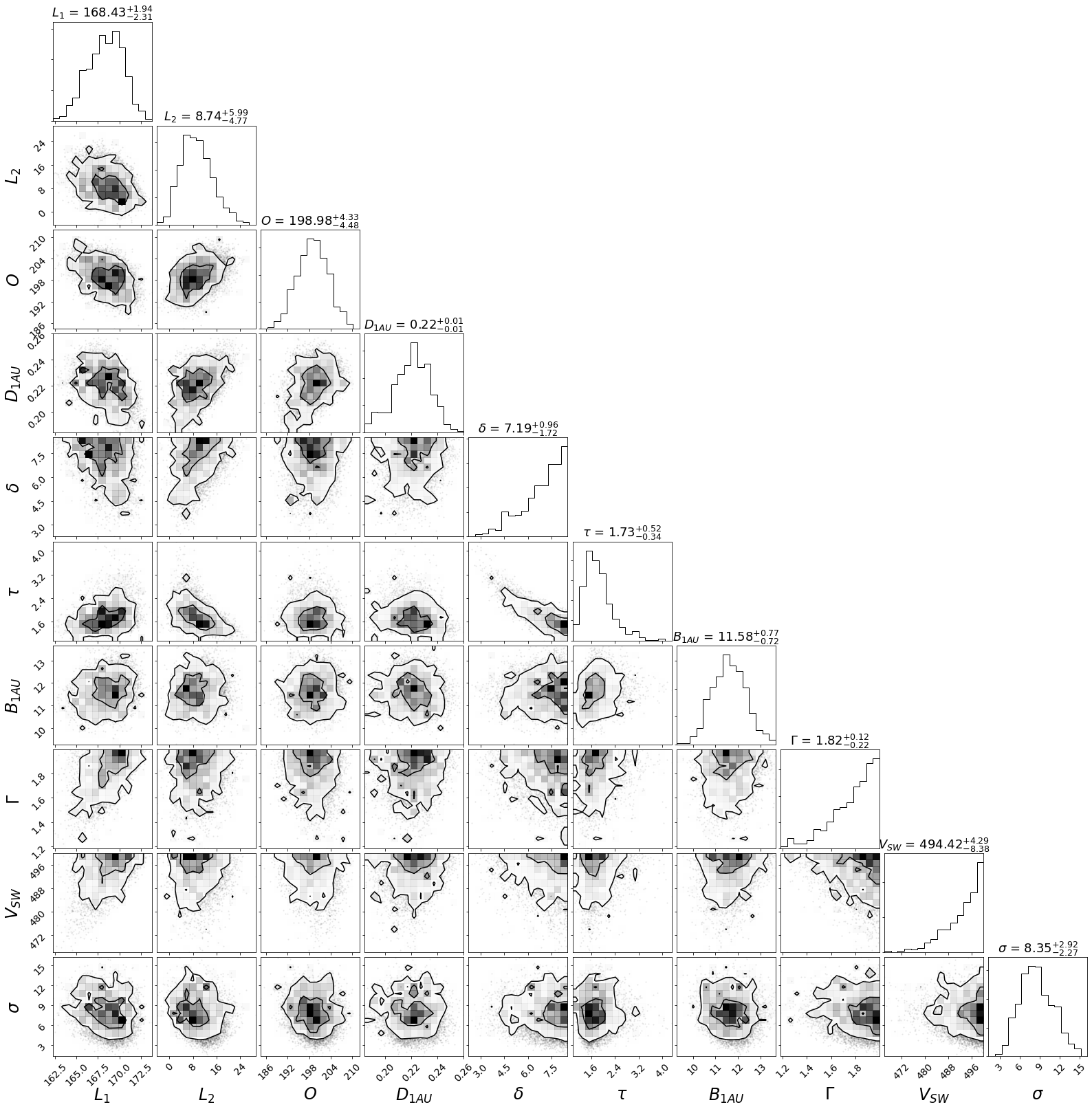}
\caption{\label{fig:40_params} Scatter-plot matrix of the results of the ABC-SMC algorithm using the PSP measurements from Figure \ref{fig:40_event}. The setup is the same as in Figure \ref{fig:30_synthetic_params} except that we now show more rows/columns as fewer parameters are fixed in the analysis. In comparison to the results from our synthetic test analysis the uncertainties on all parameters are significantly larger. We furthermore see that 3 parameters hit the boundaries that are set by our prior choice, i.e. the $\delta$, $\Gamma$ and $V_\text{SW}$ parameter all appear to be larger in our analysis than is allowed by our uniform priors. In the case of the $\delta$ parameter this is likely due to a strong bias towards larger CME's that is discussed in more detail later. The result of $V_\text{SW}$ that indicates a larger solar wind speed than is measured at any time in situ indicates that the initial CME speed of 280~km~s$^{-1}$ is being underestimated.}
\end{figure*}

\textbf{Figure \ref{fig:40_fitting}} shows the resulting ensemble fit of the magnetic field measurements. We find that we are able to generally reproduce the measurements very well within the first half of the flux rope until 04:00 UTC. Beyond that time the fit starts to diverge, especially the $B_N$ component which only reaches a maximum of $50\,$nT instead of the measured $80\,$nT. Furthermore, all members of the ensemble are significantly longer in duration and only end at 07:30 which was the defined $t_E$ marker. These issues with fitting the latter part of the flux rope are clear indicators of the distortion that is present in the measurement.

In \textbf{Figure \ref{fig:40_params}} we show the inferred 1D and 2D posterior distributions in the form of a scatter-plot matrix. The parameters that define the propagation direction and orientation of the CME are given in HCI coordinates. For most parameters the confidence intervals are similarly low as for the synthetic example, but as the flux rope fit itself still contains a significant error in addition to the distortion, these results should be treated with some suspicion. The confidence intervals on the direction and orientation parameters are lower than 10$\,\deg$. Particularly the result for the latitude $L=8\pm5\,\deg$ and the orientation parameter $O = 198\pm4\,\deg$ is interesting as they can be roughly verified using the coronagraph images from STEREO-Ahead from Figure \ref{fig:40_satplot}. By comparing our results to the coronagraph images we find that our inferred latitude is within the same region, as the structure in the coronagraph is at most propagating at an angle of $10\,\deg$ to the ecliptic. The inferred orientation parameter indicates that the structure should be tilted at an angle of around $15-20\,\deg$ to the ecliptic, which is harder to verify. This type of comparison could be made clearer with GCS results, if images of the CME existed at higher altitudes. The $D_\text{1~AU}=0.26\pm0.01~\text{AU}$ estimate is most definitely too large by at least 10-15\% due to the 1.5-hour time difference between the actual flux rope event and the fits. The $B_\text{1~AU}=11.5\pm0.8\,$nT parameter indicates that the CME was not particularly magnetically strong under the assumption that the scaling relations up to 1\,AU hold as they are defined in the model.

\begin{figure}[h]
\includegraphics[trim=200px 250px 125px 250px, clip, width=\linewidth]{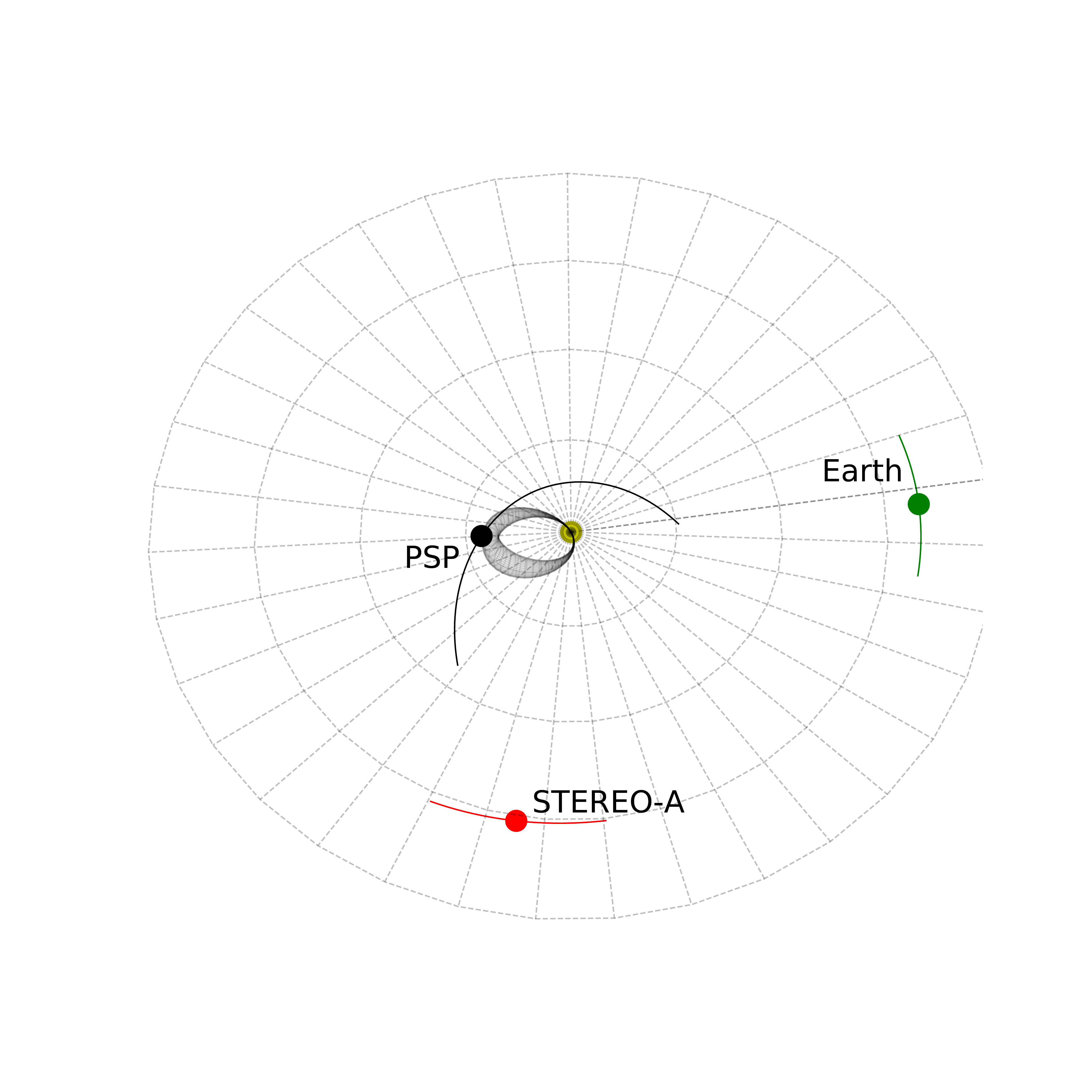}
\caption{\label{fig:40_fitting_model} Overview of the inner solar system viewing down from the solar north pole showing the Earth (green), STEREO-A (red) and PSP (black) on 2018 November 12 06:00 UT. Furthermore we show the representative 3DCORE structure as a wireframe model at the center. From this illustration we can easily see that the propagation direction (longitude) well matches the observations from the STEREO-A coronagraph.}
\end{figure}

Some of these results are put into a better perspective in \textbf{Figure \ref{fig:40_fitting_model}}, where we show the structure of the fit with respect to the planetary and satellite positions. The parameters for this single representative 3DCORE run are the medians of the ensemble solution generated by the ABC-SMC algorithm (except for the $\delta$ parameter which was set to $2$ for better visualization).

An interesting result in this particular analysis are the values for the cross-section aspect ratio $\delta$ and the magnetic field twist $\tau$. When defining the prior $p(\delta)$ we limited the maximum value to $\delta=8.5$. As we can see from Figure \ref{fig:40_params} the results prefer an extreme aspect ratio. As was already observed in more detail in Section \ref{sec:methods_synthetic} there is an inverse relationship between the aspect ratio $\delta$ and twist $\tau$. \edit1{
A large $\delta$ value is generally combined with a low magnetic field twist. The total number of twists over our torus-like shape is estimated to $\tau=1.73\pm0.5$. At 1~AU this value corresponds to around 0.6 twists/AU which is rather low. Both the aspect ratio and twist parameters are most likely the least accurate results in the analysis of this event.}
\edit1{
There is likely to be a strong bias due to the uniform priors that we have used. As our model is semi-empirical and no physical simulations are used to generate the results there is no inherent weighting for any of the parameters. This effect may be the most visible for the $D_\text{1~AU}$ and the $\delta$ parameter (and therefore indirectly also the twist). Large $D_\text{1~AU}$ and $\delta$ parameters correspond to large CME structures that therefore also have a higher chance of hitting any observer. Therefore when analyzing any event there will always be a strong bias towards larger structures as they have a large probability of having generated the measurement when assuming that small structures have the same occurrence rate as larger ones. This is most certainly not true as one would expect larger CMEs to be rarer than their smaller counterparts. This shows that in the cases where the measurements are not fully conclusive, the priors can have a strong effect on the end result.
}

The second oddity is the result for solar wind speeds $V_{SW}$ that are unusually high, with values over $470$\,km s$^{-1}$. This value is larger than any speed that is measured locally by PSP at any time just before, during or just after the CME of interest. This is a strong indicator that our chosen initial CME speed of 280 km~s$^{-1}$ is strongly underestimated. A simple explanation that can lead to such an underestimation is the viewing angle of STEREO-A. The longitudinal position of STEREO-A was approximately $230^\circ$ on November 12th (HCI), as such the viewing plane in which we measured the initial velocity was at $140^\circ$. Assuming that our inferred value of $L_1=168^\circ$ holds true nonetheless, we would have underestimated the initial speed by $12\%$ (almost 40~km~s$^{-1}$). Proper handling of this issue would require a coupling of the $L_1$ and $V_0$ parameter in our fitting algorithm \edit1{or in the model.}

\section{Conclusions \& Discussion} \label{sec:conclusion}

In this paper we have presented an improved version of the 3DCORE model and implemented an ABC-SMC fitting algorithm that we combined with our model to fit magnetic field flux rope measurements. The primary improvement with respect to the original 3DCORE model \citep{Moestl_2018} was a significant increase in the number of simulation runs that can be performed allowing the usage of a more computationally expensive fitting algorithm. This was achieved due to the definition of the geometrical model shape using a separate coordinate system which allows for efficient collision detection and evaluation of the internal analytical magnetic field. On the implementation side we made heavy use of parallelization and created a data generation pipeline that operates on multiple simulations simultaneously. In general our 3DCORE implementation is the fastest when using thousands of simulations at once.

The computational efficiency is important for two reasons. First, it could in the future open up the possibility of real-time analysis which ideally requires simulations that run on the minute-scale. Second, it will allow us to further generalize or make the model more complex without losing its usability. This will be of particular interest in future studies incorporating a more sophisticated shape or propagation model that includes a solar wind model.

We tested our ABC-SMC algorithm on a synthetic data set and found that under ideal conditions it is capable of extracting the known ground truth. This assumes that we know the exact boundaries of the flux rope signature in time and have a correct statistical model of the noise or fluctuations that occur in the measurements. This algorithm also allows us to estimate the constraints on the model parameters, which means we can determine the accuracy of our results. In the case of the synthetic test results this worked very well although we still found that there were significant errors on the cross-section aspect ratio $\delta$ and magnetic field twist $\tau$ parameters. We also found that there is a relationship between the pairs of each ($\delta$, $\tau$) solution, where a large aspect ratio corresponds to a small twist value and vice-versa. Whether this is just an artifact of our magnetic field model assumptions or if it still persists for more physically accurate models will have to be discussed in the future. If this degeneracy persists, it shows that estimating any of these two parameters independently will generally fail and auxiliary measurements will be required.

While no big issues appeared in our synthetic test case, the picture changes drastically once we apply our fitting technique to real-world data as the PSP event that we presented. The model is no longer fully capable of describing the measurements, in particular the distortion of the flux rope towards the end. The assumption of Gaussian random noise is also no longer adequate. An issue that is not apparent from the results is that the algorithm is very sensitive to how we set up the fitting points, especially the $t_S$ and $t_E$ markers. The algorithm is more efficient, in the sense of the acception-rejection ratio, for larger intervals defined by $t_S$ and $t_E$ as fewer simulations are sorted out due to generating signatures that are too long in duration. In general, setting the $t_S$ and $t_E$ markers to the exact positions of the start and end of the flux rope is impractical as the algorithm will be too slow. A larger window will lead to an over-estimation of the $D_\text{1AU}$ parameter and may also have a non-negligible effect on others.

In all our analyses we also always used uniform priors out of convenience. In the synthetic case this is no issue as the data is fully self-explanatory and can be modeled completely by our simulation. In the more complicated case, when less information can be extracted from measurements, the priors start having a significant effect on the end result. For these cases which will most likely include most observed flux ropes one has to construct more useful priors. In particular there should be a lower weighting for larger CME structures that are inherently rarer than smaller ones. Such priors could be constructed from CME catalogs that contain well behaved flux rope signatures (e.g. from HELCATS\footnote{https://www.helcats-fp7.eu/}).

The novel component of our work is the first application of Bayesian analysis with respect to CME flux ropes. For this purpose we implemented an ABC-SMC algorithm with which we can approximate the posterior distributions of all model parameters when fitting the model to observations. This approach, if done correctly, can have multiple advantages over the simpler fitting methods for flux ropes that are primarily used today. First and foremost the usage of this class of algorithms allows us to estimate the intrinsic errors  on the inferred model parameters. This gives us information on how well we are able to determine certain parameters or if they are determinable within a reasonable accuracy at all, as they may be obscured completely due to projection effects. This can furthermore allow us to compare different models and various techniques in terms of accuracy and not only in the final result, which can be ambiguous as the ground truth is unknown.

Furthermore, our fitting algorithm is easily extendable, without major changes, to the case of multi-point event analysis. In this context it is possible to investigate if and how multiple measurements agree or disagree with each other. It is additionally possible to measure, in absolute terms, the information gain that is achieved when continuously adding independent measurements to the analysis. One could, for example, perform a cost-benefits analysis on how many in situ spacecraft would be required to sufficiently determine the properties of any CME flux rope.

Lastly, the construction of priors allows us to easily incorporate additional information into our analysis. This extra data can either come from other measurements, physical considerations or previous analysis results from our algorithm. It also allows us to continuously update any result whenever new measurements or model improvements become available. 

In summary, we have developed an alternative path for fitting CME in situ flux rope measurements and presented the first results.  In the near future, we aim to further develop our technique by improving the underlying 3DCORE model, constructing a better noise model and defining better initial priors. The next logical step for our technique is to test its self-consistency when analyzing multi-point events. This will hopefully allow us to gain new insights into both our model and technique and the overall structure of CMEs.

\acknowledgments

C.M., A.J.W, R.L.B, M.A.R., T. A. thank the Austrian Science Fund (FWF): P31521-N27, P31659-N27, P31265-N27, J4160-N27. We acknowledge the NASA Parker Solar Probe Mission and FIELDS team led by S.D. Bale for use of data. We acknowledge the NASA Parker Solar Probe Mission and SWEAP team led by Justin Kasper for use of data. The open source code for the 3DCORE model is available at \url{https://github.com/ajefweiss/py3dcore}, and the \textit{heliosat} package to download and process spacecraft data is accessible at \url{https://github.com/ajefweiss/heliosat}. The python scripts that were used to generate the figures and perform analysis in this paper are partially available as Jupyter notebooks at \url{https://github.com/helioforecast/Papers}.


\bibliography{bib}

\end{document}